\documentclass[a4paper,11pt]{article}
\usepackage{jcappub}

\usepackage[T1]{fontenc} 

\title{\boldmath Symmetries of geodesic motion in G\"odel-type spacetimes}

\author[]{U. Camci}

\affiliation[]{Department of Physics, Akdeniz University, 07058 Antalya, Turkey}

\emailAdd{ucamci@akdeniz.edu.tr}

\abstract{In this paper, we study Noether gauge symmetries of geodesic motion for geodesic Lagrangian of four classes of metrics of G\"{o}del-type spacetimes for which we calculated the Noether gauge symmetries for all classes I-IV, and find the first integrals of corresponding classes to derive a complete characterization of the geodesic motion. Using the obtained expressions for $\dot{t}, \dot{r}, \dot{\phi}$ and $\dot{z}$ of each classes I-IV which depends essentially on two independent parameters $m$ and $w$, we explicitly integrated the geodesic equations of motion for the corresponding G\"{o}del-type spacetimes.   }


\begin{document}
\maketitle
\flushbottom

\section{Introduction}
\label{INT}

In 1949, Kurt G\"{o}del \cite{godel} derived an exact cosmological solution of Einstein's field equations, in which the rotation of a homogeneous mass distribution around every point is presented with a constant rotation rate. The G\"{o}del's metric is the best known example of causality violated universe model \cite{kramer}. The existence of closed timelike curves (CTCs) is particular property of G\"{o}del's universe. The rotational symmetry of G\"{o}del's metric comes from the existence of CTCs corresponding to circular orbits in specific coordinates, as pointed out by G\"{o}del \cite{godel}. Furthermore these circular orbits have discussed by Raychaudhuri and Thakurta \cite{rayc}. Rebou\c{c}as and Tiomno \cite{rebo1}, and Calv\~{a}o et al \cite{calvao1} have pointed out that the causality features of the G\"{o}del-type spacetimes depend on two independent parameters: $m$ and $w$. For $0 \leq m^2 < 4 w^2$, they have shown that there exists only one non-causal region. For $m^2 \geq 4 w^2$, there is no CTCs, in which the limiting case $m^2 = 4 w^2$ yields a completely causal and spacetime homogeneous G\"{o}del model; for $m^2 < 0$, there exists an infinite number of alternating causal and noncausal regions. The G\"{o}del metrics are mainly interesting for their high degree of symmetry \cite{rebo1,teix,rebo2,rebo3}. All classes of G\"{o}del-type spacetimes admit at least a $G_5$ group of motions. In a special case $m^2 = 4 w^2$, it has been shown that the group of motions is $G_7$, a maximal symmetry group of G\"{o}del-type spacetimes \cite{teix}.

Applying the method of effective potential to the Schwarzchild and Kerr metrics the qualitative features of their geodesics have been explored \cite{adler,misner}. The geodesic equations of motion for the general cylindrically symmetric stationary spacetimes together with their Dirac's constraint analysis and symplectic structure have been obtained, and integrated in Ref. \cite{ugur}. The geodesic equations of motion in G\"odel-type spacetimes have been analyzed by several authors. The geodesic equations for G\"odel's metric was firstly solved by Kundt \cite{kundt}, who used the Killing vectors and corresponding constants of motion. Chandrasekhar and Wright \cite{chandra} have presented an independent derivation of the solution for the geodesic equations of G\"odel's metric. Novello \emph{et.al.} \cite{novello} have provided a detailed discussion on geodesic motion in the original G\"{o}del's universe using the method of effective potential as well as the analytical solution. Rebou\c{c}as and Tiomno \cite{rebo4} have integrated the geodesic equations for the special case $m^2  = 4 w^2$ with seven isometries, where the spacetime is conformally flat and Petrov type O.  Paiva \emph{et. al.} \cite{paiva} have examined the geodesics of the Som-Raychaudhuri spacetime \cite{som}. Calv\~{a}o \emph{et. al.} \cite{calvao2} followed Novello \emph{et. al.} \cite{novello} and give a complete discussion of timelike geodesics and also treat null geodesics for G\"{o}del-type spacetimes. Grave \emph{et. al.} \cite{grave} derived the analytical solution of the geodesic equations of G\"{o}del's universe for both particles and light in a special set of coordinates. They have generalized the work of Kajari \emph{et. al.} \cite{kajari} on the solution of lightlike geodesic equations. Recently, Dautcourt \cite{daut} considered only the lightlike case, and studied the lightcone of the G\"{o}del-type metrics. Some spacetime symmetry properties of the original G\"{o}del metric and G\"{o}del-type spacetimes (see Refs.\cite{katzin}-\cite{cmc-sharif}) will be discussed in the following section.

In order to solve the geodesic equations of motion, the central idea is to find simple expressions for constants of motion, i.e. conservation laws. To derive the equations of geodesic motion, one can use the Lagrangian formalism. The Noether symmetries are interesting symmetries associated with differential equations possessing a Lagrangian, and they describe physical features of differential equations in terms of conservation laws admitted by them.  The strict Noether symmetry approach which does not include a gauge term (see Refs.\cite{capo93}-\cite{sharif2013}) is a kind of symmetry in which the Lie derivative of Lagrangian arising from the metric of interest dragging along a vector field $\bf{Y}$ vanishes, i.e. $\pounds_{\bf Y} L = 0.$ The  Noether gauge symmetry (NGS) approach \cite{feroze1}-\cite{ibrar}, as a generalization of the former strict Noether symmetry approach, will be discussed in section \ref{ngs}. The connection between the KVs and NGSs of spacetimes is examined by several works. For the spaces of different curvatures such as  maximally symmetric spacetimes and Bertotti-Robinson like spacetime, the existence of new conserved quantities has been discussed and conjectured \cite{feroze1,feroze2,feroze3}. The NGSs of FRW spacetimes have been studied by Tsamparlis and Paliathanasis \cite{tsamparlis1}. They have also examined the NGSs of class A Bianchi type homogeneous spacetimes with a scalar field that is minimally coupled to the gravity \cite{tsamparlis2}. Recently, Ali and Feroze \cite{feroze4} have provided a classification of plane symmetric static spacetimes according to their geodesic Lagrangian considering NGS approach.

Physically, the presence of conserved quantities which is directly related with Noether symmetries gives selection rule to recover classical behaviour in cosmic evolution, and they reduce the number of dynamical variables of the system due to the cyclic variables appeared. Basic geometrical symmetries, namely Lie point and Noether ones connected to the spacetimes like FRW and Bianchi type universe models have been discussed in context of $f(R), f(T)$ and scalar-tensor gravity theories (see \cite{capo93}-\cite{sharif2013} and \cite{yusuf}). Furthermore, the existence of Noether symmetries yields a classification of point singularities, where the symmetry is broken, for cosmologies coming from the extended theories of gravity (see the review of \cite{capo2012}). In the cosmological contexts, the Noether symmetry technique can play a crucial role. For example, in any gravity theory including a scalar field, the Noether symmetry gives us a specific form of coupling function and the potential \cite{camci2,jamil, ibrar}. An easy way for explaining the accelerated expansion of the universe is usually to consider the dark energy with negative pressure, and the simplest dark energy candidate takes place of the cosmological constant. Recently, it is proposed to use the Noether symmetry approach to probe the nature of dark energy \cite{basilakos}.

This study is designed as follows. In the following section, we give a short review about G\"{o}del-type spacetimes and their properties. In section \ref{eqgm}, we present the equations of geodesic motion for G\"{o}del-type spacetimes and their geodesic Lagrangian and Hamiltonian structure. In section \ref{ngs}, we explicitly discuss the Noether gauge symmetry approach of geodesic Lagrangian for G\"{o}del-type spacetimes while in section \ref{soln}, we give solution of Noether gauge symmetry equations in detail. Finally, our conclusion with a brief summary and discussions of finding is presented in Section \ref{conc}.

\section{G\"odel-type Spacetimes}
\label{fR}

In cylindrical coordinates, $x^a = (t,r,\phi,z), a = 0,1,2,3$, the line element for the G\"{o}del-type spacetimes can be written \cite{kramer,rayc}
\begin{equation}
ds^2 = \left[ dt + H(r) d\phi \right]^2 -dr^2 -D^2 (r) d\phi^2 -dz^2. \label{godel}
\end{equation}
The necessary and sufficient conditions for a G\"odel-type manifold to be spacetime
homogeneous (STH, hereafter) are found as \cite{rebo1,calvao1,teix,rebo2}
\begin{eqnarray}
& & \frac{D''}{D} = const \equiv m^2 , \label{cond1} \\& &
\frac{H'}{D} = const \equiv - 2\omega \label{cond2}
\end{eqnarray}
where prime denotes derivative with respect to the radial coordinate $r$. All STH Riemannian manifolds endowed with a G\"{o}del-type spacetime (\ref{godel}) are obtained as follows:
\\
{\bf Class I} : $m^2 > 0, \omega  \neq 0$. For this case, the
general solution of Eqs. (\ref{cond1}) and (\ref{cond2}) can be
written by
\begin{equation}
H(r) = \frac{2 \omega}{m^2} \left[ 1 - cosh(mr) \right] \quad and
\quad D(r) = \frac{1}{m} sinh(mr). \label{class1}
\end{equation}
{\bf Class II} : $m^2 = 0, \omega \neq 0$. The general solution of Eqs. (\ref{cond1}) and (\ref{cond2}) is
\begin{equation}
H(r) = - \omega r^2 \quad and \quad D(r) = r, \label{class2}
\end{equation}
where only the essential parameter $\omega$ appears.
\\
{\bf Class III} : $m^2 \equiv - \mu^2 < 0, \omega \neq 0$. Similarly, the integration of the conditions for homogenity Eqs. (\ref{cond1}) and (\ref{cond2}) leads to
\begin{equation} H(r) =
\frac{2 \omega}{\mu^2} \left[ cos(\mu r) -1 \right] \quad and
\quad D(r) = \frac{1}{\mu} sin(\mu r). \label{class3}
\end{equation}
{\bf Class IV} : $m^2 \neq 0, \omega = 0$. We refer to the
manifolds of this class as degenerated G\"odel-type manifolds,
since the cross term in the line element, related to the rotation
$\omega$ in the G\"odel model, vanishes. By a trivial coordinate
transformation, one can make $H = 0$ with $D(r)$ given, respectively, by Eqs. (\ref{class1}) or (\ref{class3}) depending on whether $m^2 > 0$ or $m^2 \equiv - \mu^2 < 0$. Throughout this paper we have used the following property
\begin{equation}\label{ozel1}
D^2 \left( \frac{D'}{D}  \right)' = -1,
\end{equation}
which is valid for STH G\"{o}del-type spacetimes only. The form of the fully-covariant Riemann curvature tensor, Weyl tensor and Ricci tensor for the G\"{o}del-type spacetime, with non-vanishing components are as follows
\begin{eqnarray}
& & R_{0101} = w^2, \,\, R_{0202} = w^2 D^2,  \,\, R_{0112} =-w^2 H^2, \,\, R_{1212} = w^2 H^2 + \left( \frac{3 w^2}{4} -m^2 \right) D^2 , \label{rie12} \\
& & C_{0101} = \frac{1}{6} (m^2 -4 w^2), \quad  C_{0202} = D^2 C_{0101},  \quad C_{0303} =-2 C_{0101}, \nonumber \\& & C_{0112} =-H C_{0101}, \quad \,\, C_{1212} = (H^2 + 2 D^2) C_{0101}, \quad C_{1313} =- C_{0101}, \label{weyl12} \\& & C_{2303} = -2 H C_{0101}, \quad C_{2323} = (2H^2 + D^2) C_{0101}  \nonumber \\
& & R_{00} = 2 w^2, \,\, R_{11} = m^2 - 2w^2 , \,\, R_{02} =  -2 w^2 H, \,\, R_{22} = m^2 D^2 - 2 w^2 (H^2 + D^2 ), \label{ricci00-22}
\end{eqnarray}
Thus the scalar curvature $R$ becomes $R = 2 (w^2 - m^2)$. The results in Refs. \cite{rayc,rebo1,teix,rebo2} for G\"odel-type manifolds can be
collected together as follows :

{\bf Theorem 1 }: The necessary and sufficient conditions for a
four-dimensional Riemannian G\"odel-type manifold to be locally
homogeneous are those given by Eqs. (\ref{cond1}) and
(\ref{cond2}).

{\bf Theorem 2} : The four-dimensional homogeneous Riemannian
G\"odel-type manifolds are locally characterized by two
independent parameters $m^2$ and $\omega$: the pair of ($m^2,
\omega$) identically specify locally equivalent manifolds.

Now let us recall the spacetime symmetries. If ${\bf X}$ be any global smooth vector field and $g_{ab}$ the metric tensor field of any type on manifold ${\it M}$, then the natural concept of a symmetry is geometrically given as a mapping, and reduced to a differential relation between ${\bf X}$ and $g_{ab}$ as \cite{katzin}
\begin{equation} \pounds_{\bf X} g_{ab} = 2 \psi g_{ab}
\label{killing}
\end{equation}
where $\pounds_{\bf X}$ is the Lie derivative operator along the
vector field ${\bf X}$, and $\psi = \psi (x^a)$ is a conformal factor.
The group of conformal motions generated by a {\it conformal Killing vector} (CKV) field ${\bf X}$ is defined by Eq. (\ref{killing}). For $\psi_{;ab} \neq 0$, the CKV field ${\bf X}$ is said to be {\it proper}, otherwise it is a special conformal Killing vector (SCKV) field ($\psi_{;ab} = 0$). The vector field ${\bf X}$ is a homothetic vector (HV) for $\psi_{,a} = 0$, and it is an isometry or a Killing vector (KV) field for $\psi = 0$. The set of all CKV (respectively SCKV, HKV and KV) form a finite-dimensional Lie algebra denoted by $\mathcal{C}$ (respectively $\mathcal{S}, \mathcal{H}$ and $\mathcal{G}$).

The KV fields as well as corresponding Lie algebra of the classes I-IV of STH G\"{o}del-type spacetimes (\ref{godel}) have been determined by Rebou\c{c}as \emph{et al.} \cite{rebo3}, and the results are stated in the following theorem:

{\bf Theorem 3} : The four-dimensional homogeneous Riemannian
G\"odel-type manifolds admit a group of isometry $G_r$ with

{\bf (i)} $r = 5$ in classes I (where $m^2 \neq 4 w^2$) , II and III;

{\bf (ii)} $r = 6$ in class IV;

{\bf (iii)} $r = 7$ in the special case of class I, where $m^2 = 4 \omega^2$.
\\
It has been pointed out that the original G\"odel metric does not admit HVs \cite{hall-costa}, of which the result is subsequently extended to the STH  G\"odel-type spacetimes \cite{melfo}. The proper CKVs and complete conformal algebra of a G\"odel-type spacetime have been computed in Ref. \cite{tsamparlis}. The Ricci collineations (RCs) and contracted RCs of STH G\"odel-type spacetimes are studied by Melfo {\it et al.} \cite{melfo}. The matter collineations (MCs) of that spacetimes are obtained by Camci and Sharif \cite{cmc-sharif}. In this work, we aim to give a complete classification for STH G\"{o}del-type spacetimes according to the Noether gauge symmetries of their geodesic Lagrangian.

\section{The Equations of Geodesic Motion} \label{eqgm}

For any spacetime metric, the freely moving massive particles or the propagation of light rays is described by geodesic equations of motion
\begin{equation}
\ddot{x}^a + \Gamma^a_{\,bc} \dot{x}^b \dot{x}^c = F^a  \label{geodesic-eq1}
\end{equation}
with the following constraint to be fulfilled
\begin{equation}
g_{ab} \dot{x}^a \dot{x}^b = \epsilon,  \label{geodesic-ceq}
\end{equation}
where the functions $\Gamma^a_{\,bc} (x^e)$ are the connection coefficients of the metric, a dot over a symbol represents  a derivative with respect to proper time $\tau$ (for massive particle motion) or with regard to an affine parameter $\lambda$ (for lightlike geodesics) along the solution curve, and $F^a (x^e)$ is the conservative force field. Here, we can write the force field as $F^a (x^e) = g^{ab} V_{,b}$, where $V(x^e)$ is the potential function. We have $\epsilon = -1,0,+1$ for spacelike, lightlike (or null) and timelike geodesics, respectively.

Using the G\"{o}del-type spacetime (\ref{godel}), a point-like Lagrangian density takes
such a form
\begin{equation}
L =  \frac{1}{2} \left[ \dot{t}^2 - \dot{r}^2 - \dot{z}^2 + (H^2(r) -D^2 (r)) \dot{\phi}^2 \right] + H(r) \dot{t} \dot{\phi} - V(t,r,\phi,z).
\label{lagr-godel}
\end{equation}
One may obtain the Euler-Lagrange equations of motion by varying of the Lagrangian (\ref{lagr-godel}) with respect to the coordinates $t, r, \phi$ and $z$ as given by
\begin{eqnarray} &&  \ddot{t} + H \ddot{\phi} + H' \dot{r} \dot{\phi} +  V_{,t} = 0,  \label{lie1}
\\ && \ddot{r} + \left( H H' -D D' \right) \dot{\phi}^2 + H' \dot{r} \dot{\phi} -V_{,r} = 0, \label{lie2}
\\ &&  H \ddot{t} + (H^2 -D^2) \ddot{\phi} + 2 (H H' -D D') \dot{r} \dot{\phi} + H' \dot{r} \dot{t} - V_{,\phi} = 0,  \label{lie3} \\ && \ddot{z}  - V_{,z} = 0. \label{lie4}
\end{eqnarray}
The \emph{energy functional} or \emph{Hamiltonian of the dynamical
system}, $E_{L}$, associated with the Lagrangian (\ref{lagr-godel}) is
found as
\begin{eqnarray}
E_{L} &=&  \dot{t} \frac{\partial L}{\partial \dot{t}}  + \dot{r}
\frac{\partial L}{\partial \dot{r}}  +
\dot{\phi} \frac{\partial L}{\partial \dot{\phi}} + \dot{z} \frac{\partial L}{\partial \dot{z}} - L \nonumber \\
&=& \frac{1}{2} \left[ \dot{t}^2 - \dot{r}^2 - \dot{z}^2 + (H^2 -D^2) \dot{\phi}^2 \right] + H \dot{t} \dot{\phi} + V(t,r,\phi,z). \label{E-L}
\end{eqnarray}
Introducing the momenta $p_a = g_{ab} \dot{x}^b = \frac{\partial L}{\partial \dot{x}^a}$ we have
$p_t = \dot{t} + H \dot{\phi}$,\,\, $p_r = -\dot{r}$,\,\, $p_{\phi}= H \dot{t}+ (H^2 -D^2)\dot{\phi}$, \,\, $p_z= -\dot{z}$. Then $E_L$ becomes
\begin{eqnarray}
E_{L} &=& \dot{x}^a p_a -  L
\nonumber \\ &=& \frac{1}{2} \left[ p_t^2 - p_{r}^2 - p_{z}^2 - \frac{1}{D^2} \left( p_{\phi} - H p_t \right)^2 \right] + V(t,r,\phi,z). \label{E-L2}
\end{eqnarray}

\section{The Noether Symmetry Approach} \label{ngs}

The {\it Noether gauge symmetry} (NGS) is defined as follows. Let us consider a vector field
\begin{equation}\label{vecf}
{\bf Y} = \xi \frac{\partial}{\partial \tau}+\eta^0
\frac{\partial}{\partial t} + \eta^1 \frac{\partial}{\partial r} +
\eta^2 \frac{\partial}{\partial \phi} + \eta^3 \frac{\partial}{\partial z}
\end{equation}
where $\xi, \eta^1, \eta^2, \eta^3$ and $\eta^4$ are depend on $\tau, t,r,\phi$
and $z$. Here, $\tau$ is an independent variable, $t(\tau), r(\tau), \phi(\tau)$
and $z(\tau)$ are dependent variables, i.e. the configuration
space of the Lagrangian (\ref{lagr-godel}) is $Q = (t, r, \phi,z)$,
whose tangent space is $TQ = (t,r,\phi,z,\dot{t},\dot{r},\dot{\phi},\dot{z})$. The first extension of the above vector field is given by
\begin{equation}\label{prol}
{\bf Y^{[1]}} = {\bf Y}+\eta^0_{\tau} \frac{\partial}{\partial
\dot{t}} + \eta^1_{\tau} \frac{\partial}{\partial \dot{r}} + \eta^2_{\tau}
\frac{\partial}{\partial \dot{\phi}} +\eta^3_{\tau}
\frac{\partial}{\partial \dot{z}},
\end{equation}
in which
\begin{equation}\label{def}
{\eta^0_{\tau}} = D_{\tau}\eta^0 - \dot{t} D_{\tau} \xi, \quad {\eta^1_{\tau}} =
D_{\tau}\eta^1-\dot{r}D_{\tau}\xi, \quad  \eta^2_{\tau} =
D_{\tau}\eta^2-\dot{\phi}D_{\tau}\xi, \quad \eta^3_{\tau} =
D_{\tau}\eta^3-\dot{z}D_{\tau}\xi,
\end{equation}
where $D_{\tau}$ is the operator of total differentiation with respect to $\tau$
\begin{equation}\label{totd}
{D_{\tau}} =  \frac{\partial}{\partial \tau} + \dot{t}
\frac{\partial}{\partial t} + \dot{r} \frac{\partial}{\partial r}+
\dot{\phi} \frac{\partial}{\partial \, \phi}+ \dot{z} \frac{\partial}{\partial z}.
\end{equation}

The vector field ${\bf Y}$ is a NGS of a Lagrangian $L(\tau,t,r,\phi,z,\dot{t},\dot{r}, \dot{\phi},\dot{z})$ if there exists a gauge function $f(\tau,t,r,\phi,z)$ such that
\begin{equation}\label{Noether}
{\bf Y}^{[1]}L + L \, (D_{\tau} \xi) = D_{\tau} f
\end{equation}
which takes the alternative form \cite{tsamparlis1,tsamparlis2}
\begin{eqnarray}
& & \xi_{,a} = 0, \qquad g_{ab} \eta^b_{,\tau} = f_{,a} \qquad  \pounds_{\bf \eta} g_{ab} = \xi_{,\tau} g_{ab},   \qquad  \pounds_{\bf \eta} V = - \xi_{,\tau} V - f_{,\tau}  \label{neq-1234}
\end{eqnarray}
where $\pounds_{\bf \eta}$ is the Lie derivative operator along ${\bf \eta} = \eta^0 \partial_t + \eta^1 \partial_r + \eta^2 \partial_{\phi}+ \eta^3 \partial_z$. It is noted here that the set of all NGSs form a finite dimensional Lie algebra denoted by $\mathcal{N}$.

The significance of NGS is clearly comes from the fact that if ${\bf Y}$ is the Noether gauge symmetry corresponding to the Lagrangian $L(\tau, x^a,\dot{x}^a)$, then
\begin{equation}\label{frstI}
{I} = \xi L + \left(\eta^a-\xi \dot{x}^a \right) \frac{\partial
L}{\partial \dot{x}^a} - f
\end{equation}
is a \emph{first integral} or a \emph{conserved quantity} associated with NGS vector field ${\bf
Y}$ \cite{noether}. Then it follows from this relation for the geodesic Lagrangian (\ref{lagr-godel}) that
\begin{equation}\label{frstI-2}
{I} = - \xi E_L +(\eta^0 + H \eta^2 ) \dot{t} - \eta^1 \dot{r} + \left[ H \eta^0 + (H^2- D^2)\eta^2 \right] \dot{\phi} - \eta^3 \dot{z} - f,
\end{equation}
where $E_L$ is given in (\ref{E-L}). Now we seek the condition in order that the Lagrangian (\ref{lagr-godel}) would admit NGS.

Recently the Noether gauge symmetries of geodesic Lagrangian for some spacetimes have been calculated, and classified according to their symmetry generators \cite{feroze1}-\cite{tsamparlis2}. Here, we investigate the NGSs of the G\"{o}del-type spacetimes. For the G\"odel-type spacetimes (\ref{godel}), the Noether gauge symmetry equations (\ref{Noether}) or (\ref{neq-1234}) yield {\it 19} partial differential equations:
\begin{eqnarray}
& & \xi_{,t} = 0, \quad \xi_{,r} = 0, \quad \xi_{,\phi} = 0, \quad \xi_{,z} = 0, \label{ngseq-1} \\& & 2 T_{,t} - \xi_{,\tau} = 0, \quad T_{,z} -Z_{,t} = 0, \quad 2 R_{,r} - \xi_{,\tau} = 0, \label{ngseq-2} \\ & & Z_{,r} + R_{,z} =0, \quad 2 Z_{,z} - \xi_{,\tau} = 0, \label{ngseq-3} \\& & T_{,r} - R_{,t} - \frac{H'}{D} P = 0,  \label{ngseq-4} \\& & Z_{,\phi} + D P_{,z} - H \, Z_{,t} = 0, \label{ngseq-5}\\& & T_{,\phi} + H' R - D P_{,t} - \frac{H}{2} \xi_{,\tau} = 0, \label{ngseq-6} \\& & R_{,\phi} - H R_{,t} - D' P + D P_{,r} = 0, \label{ngseq-7} \\& &  P_{,\phi} - H P_{,t} + D' R - \frac{D}{2} \xi_{,\tau} = 0, \label{ngseq-8} \\& &  T \, V_{,t} + R \, V_{,r} + \frac{1}{D} (V_{,\phi}- H V_{,t}) P + Z \, V_{,z} + \xi_{,\tau} V + f_{,\tau}= 0, \label{ngseq-9} \\& & T_{,\tau} - f_{,t} =0, \quad R_{,\tau} + f_{,r} = 0, \quad D P_{,\tau} + f_{,\phi} - H f_{,t} = 0, \quad Z_{,\tau} + f_{,z} = 0, \label{ngseq-10}
\end{eqnarray}
where the subscripts with comma denotes partial derivatives, and we have used the following definition
\begin{eqnarray}
& & \eta^0 = T - \frac{H}{D} P, \qquad \eta^1 = R, \qquad \eta^2 = \frac{P}{D}, \qquad \eta^3 = Z. \label{PRTZ}
\end{eqnarray}
The general solution to the above NGS equations is introduced in the next section for each classes I-IV for the G\"{o}del-type spacetimes.

\section{The Solution of Noether Symmetry Equations } \label{soln}

After some algebra, we have calculated the general solution to Eqs. (\ref{ngseq-1})-(\ref{ngseq-10}) in order to get NGSs for each of the classes  I, II, III and IV when the potential function $V(x^e)$ vanishes. From the first set of equations (\ref{ngseq-1}) we have $\xi = \xi (\tau)$. From (\ref{ngseq-2}) and (\ref{ngseq-3}) one obtains
\begin{eqnarray}
& & T = z \left[-r h_1 (\tau,\phi) + h_2 (\tau,\phi) \right] + f_1 (\tau,r,\phi) + \frac{t}{2} \xi(\tau)_{,\tau} \label{T} \\ & & R =  z \left[ t h_1 (\tau,\phi) + h_3 (\tau,\phi) \right] + g_1 (\tau,t,\phi) + \frac{r}{2} \xi(\tau)_{,\tau} \label{R}  \\& & Z = t \left[-r h_1 (\tau,\phi) + h_2 (\tau,\phi) \right] - r h_3 (\tau,\phi) + h_4 (\tau,\phi) + \frac{z}{2} \xi(\tau)_{,\tau} \label{Z}
\end{eqnarray}
where $H'\neq 0$, and $h_1 (\tau,\phi), h_2 (\tau,\phi), h_3 (\tau,\phi), h_4 (\tau,\phi), f_1 (\tau,r,\phi), g_1 (\tau,t,\phi)$ are arbitrary integration functions of their arguments. These solutions do not depend on the metric functions $D(r)$ and $H(r)$, and so they are the general solution of Eqs. (\ref{ngseq-2})-(\ref{ngseq-3}).
The equations (\ref{T})-(\ref{Z}) are the general solution to the NGS equations (\ref{ngseq-1})-(\ref{ngseq-3}) for all STH G\"{o}del-type metrics. These general solutions depend on six arbitrary functions, which will be determined by the remaining NGS equations for each different class of STH G\"{o}del-type Riemannian manifolds.

\subsection{Classes I, II and III}

Now let us try to find remaining arbitrary function $P(\tau,t,r,\phi,z)$. Using the property $H'=-2 w D$ for STH G\"{o}del-type spacetimes then the Eq.(\ref{ngseq-4}) yields
\begin{eqnarray}
& & P = \frac{1}{2 w} \left[ g_{1,t} -f_{1,r} + 2 z h_1 (\tau,\phi) \right]. \label{P}
\end{eqnarray}
Inserting (\ref{Z}) and (\ref{P}) into (\ref{ngseq-5}) one obtains $h_1 = h_2 =0, h_3 = h_3 (\tau)$ and $h_4 = h_4 (\tau)$. Thus substitution of (\ref{T}), (\ref{R}) and (\ref{P}) into the Eq. (\ref{ngseq-6}) yields $h_3 =0$, and
\begin{eqnarray}
& & f_1 = 2w \left[ D \ell_1 (\tau,\phi) + k_1 (\tau,r) \right] + \phi \left( w r D + \frac{H}{2}\right)\xi(\tau)_{,\tau} \label{f1} \\& & g_1 = \ell_{1,\phi} + g (\tau,\phi) \sin (2 w t) + h(\tau,\phi) \cos (2 w t), \label{g1}
\end{eqnarray}
where $\ell_1, k_1, g$ and $h$ are the arbitrary functions of their arguments. For vanishing potential, the Eq. (\ref{ngseq-9}) reduces to $f_{,\tau} = 0$, i.e. the gauge function is not depend on $\tau$. Using (\ref{T}) in the first equation of (\ref{ngseq-10}) we find that the component $\xi(\tau)$, the gauge function $f$, and the functions $\ell_1$ and $k_1$ have the following forms
\begin{eqnarray}
& & \xi = c_1 + a_1 \tau + a_2 \frac{\tau^2}{2}, \label{xi} \\& & f = a_2 \frac{t^2}{4} + t \left[ 2w (D L_1 (\phi) + K_1 (r)) + a_2 \phi \left( w r D + \frac{H}{2}\right) \right] + f_2 (r,\phi,z), \label{f} \\& &
\ell_1 = \tau L_1 (\phi) + L_2(\phi), \qquad k_1 = \tau K_1 (r) + K_2 (r), \label{l1-k1}
\end{eqnarray}
where $a_1, a_2, c_1$ are integration constants, and $L_1(\phi), L_2(\phi), K_1(r), K_2(r), f_2(r,\phi,z)$ are integration functions. The functions $T, R, Z$ and $P$ can now be simplified to give
\begin{eqnarray}
T  &=& 2w \left[ D (\tau L_1 + L_2) + \tau K_1 + K_2 \right] + (a_1 + a_2 \tau) \left[ \phi \left( w r D + \frac{H}{2} \right) + \frac{t}{2} \right], \label{T2} \\
R  &=&  \tau L_{1,\phi} + L_{2,\phi} + g (\tau,\phi) \sin (2w t) + h(\tau,\phi) \cos (2w t) + \frac{r}{2} (a_1 + a_2 \tau), \label{R2}\\  Z &=& h_4 (\tau) + \frac{z}{2} (a_1 + a_2 \tau), \label{Z2} \\  P &=& g (\tau,\phi) \cos (2w t) - h(\tau,\phi) \sin (2w t) - D' \left( \tau L_1 + L_2 \right) - \tau K_{1,r} - K_{2,r} \nonumber \\& & \quad - \phi \frac{r D'}{2}(a_1 + a_2 \tau). \label{P2}
\end{eqnarray}
If we substitute (\ref{R2}) and (\ref{P2}) into (\ref{ngseq-7}), then the resulting equations to be satisfied  are given by
\begin{eqnarray}
&& L_{1,\phi \phi} + L_1 -D K_{1,rr} + D' K_{1,r} + a_2 \frac{\phi}{2} (r -D D') = 0,   \label{l1k1} \\ & &  L_{2,\phi \phi} + L_2 -D K_{2,rr} + D' K_{2,r} + a_1 \frac{\phi}{2} (r -D D') = 0,  \label{l2k2} \\&& g_{,\phi} + (D' + 2 w H) h = 0, \label{gh} \\& & h_{,\phi} - (D' + 2 w H) g = 0. \label{hg}
\end{eqnarray}
It can be easily seen from Eqs. (\ref{l1k1}) and (\ref{l2k2}) that for the classes I and III,  $a_1 = a_2 =0$. In class II (where $ H = -w r^2$ and $D = r$),  $a_1$ and $a_2$ are not necessarily to be zero. For classes I and III, the Eqs.(\ref{gh}) and (\ref{hg}) yield
\begin{eqnarray}
& & g_{,\phi} + \left( \frac{2w}{m} \right)^2 h = \left[ 1 - \left( \frac{2w}{m} \right)^2 \right] h = 0,  \\ & & h_{,\phi} - \left( \frac{2w}{m} \right)^2 g = \left[ 1 - \left( \frac{2w}{m} \right)^2 \right] g= 0.
\end{eqnarray}
Thus, we have two different cases of solutions for classes I and III,
\begin{eqnarray}
&& (a) \,\, g = h = 0, \quad {\rm where}\,\, m^2 \neq 4 w^2, \label{case-a}  \\
&& (b) \,\, g_{,\phi} + h = 0, \quad h_{,\phi} - g =0, \quad {\rm where} \,\, m^2 = 4 w^2.  \label{case-b}
\end{eqnarray}
In case ($b$), the functions $g$ and $h$ have the following solutions
\begin{eqnarray}
&&  g = h_5 (\tau) \cos \phi + h_6 (\tau) \sin \phi, \quad h = h_5 (\tau) \sin\phi - h_6(\tau) \cos \phi,  \label{case-b}
\end{eqnarray}
where $h_5 (\tau)$ and $h_6 (\tau)$ are integration functions.

In classes I and III, when $m^2 \neq 4 w^2$, i.e. the case ($a$), the Eqs. (\ref{l1k1}) and (\ref{l2k2}) are satisfied only if
\begin{eqnarray}
&& L_{1,\phi \phi} + L_1 = D K_{1,rr} - D' K_{1,r} = const. = q_1,   \label{l1k1-2} \\ & &  L_{2,\phi \phi} + L_2 = D K_{2,rr} - D' K_{2,r} = const. = q_2,  \label{l2k2-2}
\end{eqnarray}
where $q_1, q_2$ are separation constants. Integrating (\ref{l1k1-2}) and (\ref{l2k2-2}) we obtain
\begin{eqnarray}
&& L_1 = a_3 \cos\phi + a_4 \sin\phi + q_1, \quad K_1 = - q_1 D + a_5 \int D dr + a_6,  \label{l1k1-s} \\ & &  L_2=  c_2 \cos\phi + c_3 \sin\phi + q_2 , \quad K_2= - q_2 D + c_4 \int D dr + c_5 ,  \label{l2k2-s} \end{eqnarray}
where $a_i$'s and $c_i$'s are constant parameters. Now the Eq.(\ref{ngseq-8}) is identically satisfied for classes I and III. The remaining (i.e. second, third and fourth) Eqs. of (\ref{ngseq-10}) yields $a_3 = a_4 = a_5 = a_6 = 0, h_4 = c_6 \tau + c_7$ and $f = -c_6 z$, where $c_6,c_7$ are integration constants. It is noted here that there is no contribution of the separation constants $q_1$  and $q_2$ to the NGS vector fields. Then the NGS vector field components and the gauge function for class I with the condition $m^2 \neq w^2$ are found as
\begin{eqnarray}
& & \xi = c_1, \,\, \eta^0 = - \frac{H}{D} (c_2 \cos \phi + c_3 \sin\phi) + c_4 \frac{2w}{m} + c_5, \,\, \eta^1 = -c_2 \sin\phi + c_3 \cos\phi, \\& & \eta^2 = -\frac{D'}{D} (c_2 \cos\phi + c_3\sin\phi) - c_4 m, \quad \eta^3 = c_6 \tau + c_7, \quad f = -c_6 z,
\end{eqnarray}
which means that we have {\it seven NGSs}, i.e. the {\it five KVs} ${\bf X}_1,..., {\bf X}_5$
\begin{eqnarray}
& & {\bf X}_1 = \partial_t, \quad {\bf X}_2 = \partial_z, \quad
{\bf X}_3 = \frac{2 \omega}{m} \partial_t - m \partial_{\phi},
\nonumber
\\ & & {\bf X}_4 = - \frac{H}{D} sin\phi \partial_t + cos\phi
\partial_r - \frac{D'}{D} sin\phi \partial_{\phi}, \label{KV-I1} \\
& & {\bf X}_5 = - \frac{H}{D} cos\phi \partial_t - sin\phi
\partial_r - \frac{D'}{D} cos\phi \partial_{\phi}, \nonumber
\end{eqnarray}
and {\it two non-Killing NGSs}
\begin{eqnarray}
& & {\bf Y}_1 = \partial_{\tau}, \qquad {\bf Y}_2 = \tau \partial_z \quad {\rm with \, gauge\, term} \ f = -z.  \label{Y12}
\end{eqnarray}
The Lie algebra has the following non-vanishing commutators:
\begin{eqnarray}
& & \left[ {\bf X}_3, {\bf X}_4 \right] = - m {\bf X}_5, \quad
\left[ {\bf X}_4, {\bf X}_5 \right] = m {\bf X}_3, \quad \left[
{\bf X}_5, {\bf X}_3 \right] = - m {\bf X}_4 . \nonumber
\end{eqnarray}
It should be noticed that the expressions for all KVs are time-independent.

In class III, where $m^2 \equiv - \mu^2 <0, \mu^2 >0$ and $w \neq 0$, it follows that the KVs ${\bf X}_1, {\bf X}_2, {\bf X}_4, {\bf X}_5$ and the non-Killing NGSs ${\bf Y}_1, {\bf Y}_2$ are same form as given the above, but only ${\bf X}_3$ has the form as $(2w / \mu )\partial_t + \mu \partial_{\phi}$.

For class I the first integrals (\ref{frstI-2}) associated with ${\bf X_1},...,{\bf X_5}, {\bf Y_1} $ and ${\bf Y_2}$ are found as
\begin{eqnarray} \label{frstI-123}
& &  I_1 = \dot{t} + H \dot{\phi}, \quad I_2 = -\dot{z}, \quad I_3 = \frac{2w}{m} I_1 -m \left[ H \dot{t} + (H^2-D^2)\dot{\phi} \right],
\end{eqnarray}
\begin{eqnarray} \label{frstI-4}
& &  I_4 = -\frac{\sin\phi}{D} \left\{ H(1 + D') \dot{t}  + \left[ H^2 + (H^2 -D^2) D'\right] \dot{\phi} \right\} - \cos\phi \, \dot{r},
\end{eqnarray}
\begin{eqnarray} \label{frstI-5}
& &  I_5 =  -\frac{\cos\phi}{D} \left\{ H(1 + D') \dot{t}  + \left[ H^2 + (H^2 -D^2) D'\right] \dot{\phi} \right\} + \sin\phi \, \dot{r},
\end{eqnarray}
\begin{eqnarray} \label{frstI-67}
& &  I_6 =  -E_L, \qquad I_7 = - \tau \dot{z} + z,
\end{eqnarray}
where the $E_L$ is the Hamiltonian (\ref{E-L2}) of the dynamical
system and yields
\begin{eqnarray} \label{EL-cI}
& &  E_L = \frac{1}{2} \left\{ I_1^2 - I_2^2 - \frac{1}{D^2} \left[ H I_1 - \left( \frac{2w}{m^2}I_1 - \frac{I_3}{m} \right) \right]^2 - \dot{r}^2  \right\}.
\end{eqnarray}
Hence the Hamiltonian $E_{L}$ is conserved ($\partial_{\tau} E_{L} = \frac{d E_L}{d \tau} =0$). Therefore, the vector field ${\bf Y_1}$ is the trivial NGS. Here the constants of motion $p_t \equiv I_1,  p_{\phi} \equiv 2w I_1 /m^2 -I_3 /m$ and $p_z \equiv I_2$ represent the conservation of energy, angular momentum and $z$ component of momentum, respectively. It can be solved $\dot{t}, \dot{\phi}, \dot{r}$ and $\dot{z}$ from Eqs. (\ref{frstI-123})-(\ref{frstI-67}). Furthermore, we can see from the Eq.(\ref{geodesic-ceq}) that $\epsilon = -2 I_6$, where $\epsilon = -1, 0 +1$ for spacelike, null and timelike geodesics, respectively. Then the integration for the coordinate $z$ easily give  $z= - p_z \tau + I_7,$  and we obtain all $\dot{x}^a$ and a constraint equation from the first integrals given above as
\begin{eqnarray}
&& \dot{t} = \frac{1}{D^2} \left[ p_{\phi} H + p_t (D^2 -H^2) \right], \label{t-cI} \\ && \dot{\phi} =  \frac{1}{D^2} \left[ p_t H - p_{\phi} \right], \label{phi-cI} \\& & \dot{z} = -p_z,  \label{z-cI} \\& & \dot{r}^2 = p_t^2 - V(r), \label{r2} \\& &  \dot{r} = -I_4 \cos \phi + I_5 \sin \phi, \label{r1} \\& & p_t H + p_{\phi} D' + (I_4 \sin\phi + I_5 \cos\phi ) D =0, \label{ceq-cI}
\end{eqnarray}
where we have defined the \emph{effective potential}
\begin{eqnarray}
 & & V(r) := \frac{1}{D^2} \left( p_t H - p_{\phi} \right)^2 + p_z^2 + \epsilon. \label{effpot-cI}
\end{eqnarray}
The Eq. (\ref{r2}) with the effective potential given in (\ref{effpot-cI}) is the generalization of radial equation given in Ref. \cite{calvao2}. Furthermore, we have seen that there exist a new radial equation, the Eq. (\ref{r1}), which depends only on $\phi$, not $r$. Differentiating (\ref{r1}) with respect to $\tau$ and using $\dot{\phi}$ of (\ref{phi-cI}) and the constraint Eq. (\ref{ceq-cI}) we find the following Lienard type differential equation \cite{polyanin}
\begin{equation}\label{r3}
\ddot{r} = - \frac{1}{D^3} \left( p_t H - p_{\phi} \right) \left( p_t H + p_{\phi} D' \right).
\end{equation}
The substitution of $W(r) = \dot{r}$ leads to an Abel differential equation of the second kind as
\begin{equation}\label{r3}
W W' = - \frac{1}{D^3} \left( p_t H - p_{\phi} \right) \left( p_t H + p_{\phi} D' \right),
\end{equation}
which can be written as
\begin{equation}
\left( W^2 \right)' = - \left[ \frac{1}{D^2} \left( p_t H - p_{\phi} \right)^2  \right]', \label{r4}
\end{equation}
yielding
\begin{equation}
W^2 \equiv \dot{r}^2 = -  \frac{1}{D^2} \left( p_t H - p_{\phi} \right)^2 + r_0, \label{r5}
\end{equation}
which is explicitly equivalent to the radial equation (\ref{r2}), where $r_0 = p_t^2 - p_z^2 - \epsilon$. Introducing a new variable $u=m^2 H/4 w$ which is equivalent to $\sinh^2 (m r/2)$ for the class I (see the Eq. (42) of ref. \cite{calvao2}), the Eq. (\ref{r2}) gives
\begin{equation}\label{deq1-cI}
\dot{u}^2 = m^2 p_t^2 \left[ - \eta u^2 + (1- \beta^2 + 2 w \gamma) u - \frac{m^2}{4} \gamma^2  \right],
\end{equation}
where $\eta, \, \beta^2$ and $\gamma$ are defined as
\begin{equation}\label{beta-gama}
\eta = \beta^2 -1 + \frac{4 w^2}{m^2} \qquad \beta^2 = \frac{p_z^2 + \epsilon}{p_t^2}, \qquad \gamma = \frac{p_{\phi}}{p_t}.
\end{equation}
Using the above definitions the effective potential takes the form of
\begin{eqnarray}
 & & V(r)= \frac{p_t^2 }{D^2} \left( H - \gamma^2 \right)^2 + \beta^2 p_t^2 . \label{effpot2-cI}
\end{eqnarray}
Therefore from the radial equation (\ref{r2}), one can accomplish a complete characterization of the motion which depends essentially on the parameters $\beta, \gamma, m$ and $w$. This characterization separates the motion into three distinct cases $\gamma > 0, \gamma = 0$, and $\gamma < 0$. For the trajectories of physical particles, it follows from the Eq. (\ref{r2}) that $0 \leq \beta^2 \leq 1$ (see also Refs. \cite{calvao2,novello}).

The general solution of (\ref{deq1-cI}) is given by
\begin{equation}\label{soln-deq1-cI}
u (\tau) = \frac{1}{2 \eta} \left[ 1-  \beta^2 + 2 w \gamma + \sqrt{(1- \beta^2 + 2 w \gamma)^2-\eta m^2 \gamma^2} \sin ( m p_t \sqrt{\eta} (\tau- \tau_0) ) \right],
\end{equation}
where $\eta \neq 0$ and $(1- \beta^2 + 2 w \gamma)^2-\eta m^2 \gamma^2 \geq 0$. Using the new variable $u= m^2 H/4 w$ in Eqs. (\ref{t-cI}) and (\ref{phi-cI}) for $t(\tau)$ and $\phi (\tau)$ we find
\begin{eqnarray}
&& \dot{t} = \frac{p_t (1+ w \gamma)}{u} + p_t \left( 1- \frac{4 w^2}{m^2} \right) \frac{u}{1+ u}, \label{deq2-cI}   \\ && \dot{\phi} = \frac{w p_t}{1+ u} - \frac{p_t m^2 \gamma /4}{ u ( 1 + u)}, \label{deq3-cI}
\end{eqnarray}
Now we shall use the solution (\ref{soln-deq1-cI}) of the radial equation to solve the above equations for $t(\tau)$ and $\phi (\tau)$. For $\gamma \neq 0$, after some algebra, we have obtained the following general solutions of the Eqs.(\ref{deq2-cI}) and (\ref{deq3-cI})
\begin{eqnarray}
& & t( \tau) =  \frac{2 w (\gamma + 4 w / m^2)}{m \sqrt{\eta} \sqrt{(1+p)^2 -q^2}} \arctan \left[ \frac{(1+ p) \tan(m p_t \sqrt{\eta} (\tau- \tau_0) /2 ) + q }{\sqrt{ (1+ p)^2 -q^2}} \right] \nonumber \\& & \qquad \qquad + p_t \left( 1- \frac{4 w^2}{m^2} \right) \tau + t_0,  \\ & & \phi (\tau) = \frac{ m (\gamma + 4 w / m^2)}{2 \sqrt{\eta} \sqrt{(1+p)^2 -q^2}} \arctan \left[ \frac{(1+ p) \tan(m p_t \sqrt{\eta} (\tau- \tau_0) /2 ) + q }{\sqrt{ (1+ p)^2 -q^2}} \right] \nonumber \\& & \qquad \qquad - \arctan \left[ \frac{2 \sqrt{\eta}}{m \gamma} \left\{  p \tan (m p_t \sqrt{\eta} (\tau- \tau_0)/2 ) + q \right\} \right] + \phi_0,
\end{eqnarray}
where $(1+ p)^2 > q^2, \, t_0 = t(0)$ and $\phi_0 = \phi (0)$ are integration constants. Here we have introduced the parameters $p$ and $q$ as
\begin{equation}\label{p-q}
p:= \frac{1- \beta^2 + 2 w \gamma}{2 \eta}, \qquad q := \sqrt{p^2- \frac{m^2 \gamma^2}{4 \eta}},
\end{equation}
where $p^2 \geq m^2 \gamma^2 / 4 \eta$. For $\gamma = 0$, the general solution of the Eq. (\ref{deq1-cI}) yields
\begin{equation}\label{soln2-deq1-cI}
u (\tau) = \frac{( \beta^2 - 1)}{2 \eta} \left[ -1 + \sin ( m p_t \sqrt{\eta} (\tau- \tau_0) ) \right],
\end{equation}
and using this solution in Eqs. (\ref{deq2-cI}) and (\ref{deq3-cI}) it follows that
\begin{eqnarray}
& & t( \tau) =  \frac{8 w^2 }{m^3 \sqrt{\eta (1+2 p)}} \arctan \left[ \frac{(1+ p) \tan(m p_t \sqrt{\eta} (\tau- \tau_0) /2 ) - p }{\sqrt{ 1+ 2 p}} \right]  \nonumber \\& & \qquad \qquad + p_t \left( 1- \frac{4 w^2}{m^2} \right) \tau + t_0,  \\ & & \phi (\tau) = \frac{ 2w }{m  \sqrt{\eta (1+p)}} \arctan \left[ \frac{(1+ p) \tan(m p_t \sqrt{\eta} (\tau- \tau_0) /2 ) - p }{\sqrt{ 1+ 2 p }} \right] + \phi_0,
\end{eqnarray}
where $p = (1-\beta^2)/2 \eta$.

In the special class I case, where $m^2 = 4 \omega^2$, i.e. $m = + 2 w$, which comes from the case ($b$), Eq.(\ref{ngseq-8}) is also identically satisfied, and the remaining (i.e. second, third and fourth) Eqs. of (\ref{ngseq-10}) give $a_3 = a_4 = a_5 =a_6 =0$, $h_4 (\tau) = c_6 \tau + c_7, h_5  = c_8, h_6 = c_9,$ and $f = -c_6 z$, where $c_6, c_7, c_8, c_9$ are integration constants. Hence the quantities $P,R,T,Z$ have the form
\begin{eqnarray}
&& P = c_8 \cos (2 w t + \phi) + c_9 \sin ( 2w t + \phi) - D' (c_2 \cos \phi + c_3 \sin \phi ) - c_4 \frac{D}{2 w}, \\ & & R = c_8 \sin (2 w t + \phi) - c_9 \cos ( 2w t + \phi) - c_2 \sin \phi + c_3 \cos \phi, \\& & T = 2 w D ( c_2 \cos \phi + c_3 \sin \phi) + c_4 \int D dr + c_5, \\& & Z = c_6 \tau + c_7.
\end{eqnarray}
Thus using the definition (\ref{PRTZ}) we find
\begin{eqnarray}
& & \xi = c_1, \,\, \eta^0 = - \frac{H}{D} (c_2 \cos \phi + c_3 \sin\phi) + c_4 \frac{2w}{m} - \frac{H}{D} \left[ c_8 \cos(m t + \phi)+ c_9 \sin(mt + \phi) \right]+ c_5, \nonumber \\& & \eta^1 = -c_2 \sin\phi + c_3 \cos\phi + c_8 \sin(m t + \phi)-c_9 \cos(m t + \phi), \nonumber \\& & \eta^2 = -\frac{D'}{D} (c_2 \cos\phi + c_3\sin\phi) + \frac{1}{D} \left[ c_8 \cos(m t + \phi)+ c_9 \sin(mt + \phi) \right] - c_4 m,  \\& & \eta^3 = c_6 \tau + c_7, \quad f = -c_6 z. \nonumber
\end{eqnarray}
Then one finds that there are {\it nine NGSs} which are {\it seven KVs} ${\bf X}_1,...,{\bf X}_7$ given by
\begin{eqnarray}
& & {\bf X}_1 = \partial_t, \quad {\bf X}_2 = \partial_z, \quad
{\bf X}_3 = \partial_t - m \partial_{\phi}, \nonumber
\\ & & {\bf X}_4 = - \frac{H}{D} sin\phi \partial_t + cos\phi
\partial_r - \frac{D'}{D} sin\phi \partial_{\phi},  \nonumber \\
& & {\bf X}_5 = - \frac{H}{D} cos\phi \partial_t - sin\phi
\partial_r - \frac{D'}{D} cos\phi \partial_{\phi}, \label{KV-I2}  \\&
& {\bf X}_6 = -\frac{H}{D} cos(m t + \phi) \partial_t + sin(m t +
\phi) \partial_r + \frac{1}{D} cos(m t + \phi) \partial_{\phi},
\nonumber \\& & {\bf X}_7 = -\frac{H}{D} sin(m t + \phi)
\partial_t - cos(m t + \phi) \partial_r + \frac{1}{D} sin(m t +
\phi) \partial_{\phi}, \nonumber
\end{eqnarray}
and {\it two non-Killing NGSs}  given by (\ref{Y12}), where $ m = + 2\omega$. The corresponding Lie algebra has the following non-vanishing commutators:
\begin{eqnarray}
& & \left[ {\bf X}_3, {\bf X}_4 \right] = - m {\bf X}_5, \quad
\left[ {\bf X}_4, {\bf X}_5 \right] = m {\bf X}_3, \quad \left[
{\bf X}_5, {\bf X}_3 \right] = - m {\bf X}_4 , \nonumber \\& &
\left[ {\bf X}_1, {\bf X}_6 \right] = - m {\bf X}_7, \quad \left[
{\bf X}_6, {\bf X}_7 \right] = m {\bf X}_1, \quad \left[ {\bf
X}_1, {\bf X}_7 \right] =  m {\bf X}_6 . \nonumber
\end{eqnarray}
The first integrals (\ref{frstI-123})-(\ref{frstI-67}) for ${\bf X}_1,...,{\bf X}_5, {\bf Y}_1$ and ${\bf Y}_2$ together with $m=+ 2 w$ are same ones for this special class I. The remaining first integrals associated with ${\bf X_6}$ and ${\bf X_7}$ given in (\ref{KV-I2}) are
\begin{eqnarray} \label{frstI-8}
& &  I_8 = -\frac{\cos(m t + \phi)}{D} \left\{ \left( H - \frac{2w}{m^2} \right) I_1  + \frac{I_3}{m} \right\} - \sin (m t + \phi) \, \dot{r}, \\& &  I_9 = -\frac{\sin(m t + \phi)}{D} \left\{ \left( H - \frac{2w}{m^2} \right) I_1  + \frac{I_3}{m} \right\} + \cos (m t + \phi) \, \dot{r}.
\end{eqnarray}
which yields
\begin{eqnarray}
& & \dot{r} = - I_8 \sin (mt + \phi)  + I_9 \cos (m t + \phi), \label{r1-cI2} \\& & p_t H -p_{\phi} + \left[ I_8 \cos (mt + \phi) + I_9 \sin (m t + \phi) \right] D = 0. \label{ceq-cI2}
\end{eqnarray}

For the class II, where $H(r) = - w r^2$ and $D(r) = r$, as earlier mentioned from the Eqs. (\ref{l1k1}) and (\ref{l2k2}), the constant parameters $a_1, a_2$ are not necessarily to be zero. But in this class, the Eq. (\ref{ngseq-8}) is not identically satisfied, and gives $a_1 = a_2 = 0$ and it follows from the Eq. (\ref{ngseq-7}) that  the functions $g$ and $h$ are also vanishing. The remaining part of calculation of the NGS vector field components is similar to the case of class I. After rearranging the constant parameters, the NGS vector field components and the gauge function of class II are obtained as
\begin{eqnarray}
& & \xi = c_1, \quad \eta^0 = w r (c_2 \cos \phi + c_3 \sin\phi) + c_4, \quad \eta^1 = -c_2 \sin\phi + c_3 \cos\phi, \\& & \eta^2 = -\frac{1}{r} (c_2 \cos\phi + c_3\sin\phi) + c_5, \quad \eta^3 = c_6 \tau + c_7, \quad f = -c_6 \tau,
\end{eqnarray}
which give {\it seven NGSs}, i.e. the {\it five KVs} ${\bf X}_1,..., {\bf X}_5$ by
\begin{eqnarray}
& & {\bf X}_1 = \partial_t, \quad {\bf X}_2 = \partial_z, \quad
{\bf X}_3 = \partial_{\phi}, \nonumber
\\ & & {\bf X}_4 = - \omega \, r \, sin\phi \partial_t - cos\phi
\partial_r + \frac{1}{r} sin\phi \partial_{\phi}, \label{KV-II} \\
& & {\bf X}_5 = - \omega \, r \, cos\phi \partial_t + sin\phi
\partial_r + \frac{1}{r} cos\phi \partial_{\phi} \nonumber
\end{eqnarray}
and {\it two non-Killing NGSs} same as (\ref{Y12}). The Lie algebra will have the following non-vanishing commutators:
\begin{eqnarray}
& & \left[ {\bf X}_3, {\bf X}_4 \right] = {\bf X}_5, \quad \left[
{\bf X}_3, {\bf X}_5 \right] =  - {\bf X}_4, \quad \left[ {\bf
X}_4, {\bf X}_5 \right] = 2 \omega {\bf X}_1 . \nonumber
\end{eqnarray}
Hence, the first integrals associated with those vector fields are
\begin{eqnarray}
& &  I_1 = \dot{t} -w r^2 \dot{\phi}, \quad I_2 = -\dot{z}, \quad I_3 = -w r^2 \dot{t} + r^2 (w^2 r^2 -1)\dot{\phi}, \nonumber \\& & I_4 = \sin\phi \left[ w r I_1 - \frac{1}{r} I_3 \right] - \dot{r} \cos \phi, \nonumber \\& & I_5 = \cos\phi \left[ w r I_1 - \frac{1}{r} I_3 \right] + \dot{r} \sin \phi,  \label{frst-II} \\& & I_6 = \frac{1}{2} \left[ -\dot{t}^2 + \dot{r}^2 + r^2 (1- w^2 r^2 ) \dot{\phi}^2 + \dot{z}^2 \right] + w r^2 \dot{t} \dot{\phi}, \nonumber \\& & I_7 = - \tau \dot{z} + z, \nonumber
\end{eqnarray}
where $I_6 = - E_L$ and $\epsilon = -2 I_6$ which takes the values $-1,0$ and $1$ for spacelike, null and timelike geodesics, respectively. Using the obtained first integrals given above, it follows that the coordinate $z$ is $z (\tau) = - p_z \tau + I_7$ and
\begin{eqnarray}
& &  \dot{t} = p_t \left( 1- w^2 r^2\right) - p_{\phi} w, \label{t-cII} \\& & \dot{\phi} = -w p_t  - \frac{p_{\phi}}{r^2}, \label{phi-cII} \\& & \dot{r}^2 = p_t^2 -V(r),  \label{r2-cII} \\ & & \dot{r} = -I_4 \cos\phi + I_5 \sin\phi, \label{r1-cII} \\& & I_4 \sin\phi + I_5 \cos \phi = w p_t r - \frac{p_{\phi}}{r},  \label{ceq-cII}
\end{eqnarray}
where $p_t = I_1,\, p_z = I_2$ and $p_{\phi} = I_3$ are constants of motion related with energy, z component of momentum and angular momentum, respectively, and the effective potential is defined as \begin{equation}\label{effpot-cII}
V(r)=  \left[ w p_t r + \frac{p_{\phi}}{r} \right]^2  +p_z^2 + \epsilon.
\end{equation}
It is easily seen that one can derive the radial equation (\ref{r2-cII}) from Eqs. (\ref{phi-cII}), (\ref{r1-cII}) and (\ref{ceq-cII}). Introducing the new variable $v =r^2$ the radial Eq. (\ref{r2-cII}) becomes
\begin{equation}\label{neq-cII}
\dot{v}^2 = -4 w^2 p_t^2 v^2 + 4 \alpha v -4 p_{\phi}^2,
\end{equation}
with $\alpha$ defined by $\alpha = p_t^2 -2 w p_t p_{\phi} - p_z^2 -\epsilon$. The general solution of (\ref{neq-cII}) is
\begin{equation}\label{coz-r2-cII}
v (\tau) \equiv r^2 = \frac{\alpha}{2 w^2 p_t^2 } + \sqrt{\alpha^2 - 4 w^2 p_t^2 p_{\phi}^2} \sin \left[ 2 w p_t (\tau-\tau_0)\right],
\end{equation}
where $\alpha  \geq  \pm 2 w p_t p_{\phi}$.  Then after substitution (\ref{coz-r2-cII}) into (\ref{t-cII}) the general solution of the resulting differential equation for $t(\tau)$ gives
\begin{equation}\label{coz-t-cII}
t(\tau) = \left( p_t - w p_{\phi} - \frac{\alpha}{2 p_t} \right) \tau  + \frac{w}{2} \sqrt{\alpha^2 - 4 w^2 p_t^2p_{\phi}^2} \cos \left[ 2 w p_t (\tau-\tau_0)\right] + t_0.
\end{equation}
Finally, using (\ref{coz-r2-cII}) in (\ref{phi-cII}) the behavior of the coordinate $\phi$ is given by
\begin{equation}\label{coz-phi-cII}
\phi (\tau) = -w p_t \tau  + \frac{2 w p_t p_{\phi}}{\sqrt{4 \beta^2 w^4 p_t^4 - \alpha^2}} \tanh ^{-1} \left\{ \frac{ \alpha \tan \left[ w p_t (\tau-\tau_0)\right] + 2 \beta w^2 p_t^2}{\sqrt{4 \beta^2 w^4 p_t^4 - \alpha^2}} \right\} + \phi_0,
\end{equation}
where $\beta = \sqrt{\alpha^2 - 4 w^2 p_t^2 p_{\phi}^2}$ and $4 \beta^2 w^4 p_t^4 > \alpha^2 \geq 4 w^2 p_t^2 p_{\phi}^2$. For a special case such as $p_{\phi} = I_3 = 0$ the first integrals simplify considerably. In the latter special case the geodesic equations can be integrated completely and the solution reads
\begin{eqnarray}
& &  t(\tau) = \left[ p_t - \frac{1}{2 p_t} (I_4^2 + I_5^2 ) \right] \tau \nonumber \\& & \qquad  \quad + \frac{1}{4 w p_t^2} \left[ (I_4^2 -I_5^2) \sin( 2 w  p_t \tau) - 4 I_4 I_5 \cos^2 (w p_t \tau) \right] + t_0, \label{soln-II-t} \\& & r (\tau) = \frac{1}{w p_t} \left[ I_5 \cos (w p_t \tau) - I_4 \sin (w p_t \tau) \right], \label{soln-II-r} \\& &  \phi(\tau) = -w p_t  \tau + \phi_0, \label{soln-II-phi} \\ & & z (\tau) = - p_z \tau + I_7, \label{soln-II-z}
\end{eqnarray}
where $p_t^2 - \left( I_2^2 + I_4^2 + I_5^2 \right) = \epsilon$, and $\phi_0$ is an integration constant.

\subsection{Class IV}

In this class, where $m^2 \neq 0, w = 0$, the metric functions are taken as $H(r) = 0$ and $D(r) = \frac{1}{m} \sinh(m r)$ for $m^2 >0$, or $D(r) = \frac{1}{\mu} \sin(\mu r)$ for $\mu^2 = -m^2 >0$.
Here the functions $T, R,Z$ given by the Eqs. (\ref{T})-(\ref{Z}), which are the general solution to the NGS equations (\ref{ngseq-1})-(\ref{ngseq-3}), have the same form. For this class the equation (\ref{ngseq-4}) yields that $h_1 = 0$ and
\begin{eqnarray}
& & f_1 = r h_5 (\tau, \phi) + h_6 (\tau, \phi), \qquad g_1 = t h_5 (\tau, \phi) + h_7 (\tau, \phi),  \label{f1g1-iv}
\end{eqnarray}
where $h_5, h_6, h_7$ are integration functions of their arguments. For vanishing potential, the Eq. (\ref{ngseq-9}) and first equation of (\ref{ngseq-10}) give
\begin{eqnarray}
& & \xi = c_1 + b_1 \tau + b_2 \frac{\tau^2}{2}, \label{xi-iv} \\& & f = b_2 \frac{t^2}{4} + t \left[ z L_1 (\phi) + r L_3 (\phi) + L_5 (\phi) \right] + f_2 (r,\phi,z), \label{f-iv} \\& & h_2 = \tau L_1(\phi) + L_2 (\phi), \quad h_5 = \tau L_3(\phi) + L_4 (\phi), \quad h_6 = \tau L_5 (\phi) + L_6 (\phi)
\end{eqnarray}
where $L_i (\phi)$'s and $f_2 (r, \phi,z)$ are integration functions, $b_1$ and $b_2$ are integration constants. Thus the function $P$ follows from the Eq. (\ref{ngseq-5}) as
\begin{eqnarray}
& & P = \frac{z}{D} \left[ -t \left( \tau L_{1,\phi} + L_{2,\phi} \right) + r h_{3,\phi} - h_{4,\phi} \right] + K (\tau, t,r,\phi), \label{P-iv}
\end{eqnarray}
where $K(\tau, t,r,\phi)$ is an integration function. Using the functions $P$ and $T$ in Eq.(\ref{ngseq-6}) reads $L_1 = a_1, L_2 = a_2$ and
\begin{equation}\label{K1}
K = \frac{t}{D} \left[ r h_{5,\phi} + h_{6,\phi} \right] + f_3 (\tau, r,\phi),
\end{equation}
where $f_3$ is a function of integration, and $a_1,a_2$ are constants. Using the Eq. (\ref{ngseq-7}) one gets that $h_3$ and $h_4$ are only depend on $\tau$, and $L_3 =a_3, L_4 = a_4, L_5 = a_5, L_6 = a_6$, where $a_1,...,a_6$ are constants, and $f_3 = D' h_{7,\phi} + D h_8 (\tau,\phi)$, $h_8$ is an integration function. Inserting the obtained results into Eq. (\ref{ngseq-8}), after some algebra, we get that $a_3 = a_4 = b_1 = b_2 = 0, h_3 = 0, h_7 = k_1 (\tau) \cos\phi + k_2 (\tau) \sin \phi$ and $h_8 = k_3 (\tau)$, in which the quantities $k_1,k_2$ and $k_3$ are integration functions. Finally, the remaining three equations of (\ref{ngseq-10}) yield $a_1 = 0, k_1 = a_7, k_2 = a_8, k_3 = a_9, h_4 = a_{10} \tau + a_{11}$ and $f_2 = -a_{10} \tau + a_{12}$, where $a_7,...,a_{12}$ are constants. Putting these results into the functions $P, R, T, Z$ and $f$, they take the following form
\begin{eqnarray}
& & P = D' \left( -a_7 \sin \phi + a_8 \cos \phi \right) + a_9 D,
\\ & & R = a_7 \cos \phi + a_8 \sin\phi, \\& & T = a_2 z + a_5 \tau + a_6, \\& & Z = a_2 t a_{10} \tau + a_{11}, \quad f = a_5 t -a_{10} z.
\end{eqnarray}
Rearranging the constant parameters, the NGS vector fields and the gauge function for this class are
\begin{eqnarray}
& & \xi = c_1, \quad \eta^0 = c_2 z + c_3 s + c_4, \quad \eta^1 = c_5 \cos\phi + c_6 \sin\phi, \\& & \eta^2 = \frac{D'}{D} (- c_5 \sin\phi + c_6\cos\phi ) + c_7, \quad \eta^3 = c_2 t + c_8 \tau + c_9, \quad f = c_3 t -c_8 \tau.
\end{eqnarray}
This result yields {\it nine NGSs}, i.e., the {\it six KVs} ${\bf X}_1,..., {\bf X}_6$ by
\begin{eqnarray}
& & {\bf X}_1 = \partial_t, \quad {\bf X}_2 = \partial_z, \quad
{\bf X}_3 = z \, \partial_t + t \partial_z, \nonumber
\\ & & {\bf X}_4 = cos\phi \partial_r - \frac{D'}{D} sin\phi \partial_{\phi}, \label{KV-IV} \\
& & {\bf X}_5 = - sin\phi \partial_r - \frac{D'}{D} cos\phi
\partial_{\phi}, \quad {\bf X}_6 = \partial_{\phi}, \nonumber
\end{eqnarray}
{\it two non-Killing NGSs} same as (\ref{Y12}), and extra {\it one non-Killing NGS}
\begin{eqnarray}
& & {\bf Y}_3 = \tau \partial_t \quad {\rm with \, gauge\, term} \ f = t.  \label{Y3}
\end{eqnarray}
The non-vanishing commutators of NGSs are
\begin{eqnarray}
& & \left[ {\bf X}_1, {\bf X}_3 \right] = {\bf X}_2, \quad \left[
{\bf X}_2, {\bf X}_3 \right] = {\bf X}_1 , \nonumber \\& & \left[
{\bf X}_4, {\bf X}_5 \right] = - m^2 {\bf K}_6 , \quad \left[{\bf
X}_5, {\bf X}_6 \right] = {\bf X}_4 , \quad \left[ {\bf X}_6, {\bf
X}_4 \right] = {\bf X}_5 . \nonumber
\end{eqnarray}
If $m^2 = \omega = 0$, then the line element (\ref{godel}) is
clearly Minkowskian. Therefore, this particular case has not been
included in this study. The first integrals of this class are
\begin{eqnarray}
& &  I_1 = \dot{t}, \quad I_2 = -\dot{z}, \quad I_3 = z \dot{t} - t \dot{z}, \nonumber \\& & I_4 = -\dot{r} \cos\phi  + \dot{\phi} D D' \sin\phi  , \nonumber \\& & I_5 = \dot{r} \sin\phi  + \dot{\phi} D D' \cos\phi ,  \label{frst-IV} \\& & I_6 = - D^2 \dot{\phi}, \quad I_7 = \frac{1}{2} \left[ -\dot{t}^2 + \dot{r}^2 + D^2 \dot{\phi}^2 + \dot{z}^2 \right], \nonumber \\& & I_8 = - \tau \dot{z} + z, \quad I_9 =  \tau \dot{t} - t, \nonumber
\end{eqnarray}
where $I_7 = - E_L$. From the Eq.(\ref{geodesic-ceq}), we get that $\epsilon = - 2 I_7$, and $\epsilon = -1,0,1$ for spacelike, lightlike and timelike geodesics, respectively. The above constants of motion can be solved for $\dot{t}, \dot{r}, \dot{\phi}$, and $\dot{z}$. Then, we find from (\ref{frst-IV}) that $t = \tau I_1 - I_9$ and $z = - I_2 \tau + I_8$, which means that along any geodesic the time coordinate $t$ and the axial coordinate $z$ vary uniformly with respect to its affine parameter $\tau$. It also follows from (\ref{frst-IV}) that
\begin{eqnarray}
&& \dot{\phi} = - \frac{I_6}{D^2}, \label{phi-iv} \\
&& \dot{r}^2 = I_1^2 -I_2^2 + 2 I_7 - \frac{I_6^2}{D^2}, \label{r2-iv} \\
&& \dot{r} = - I_4 \cos\phi + I_5 \sin\phi , \label{r-iv} \\& & I_4 \sin \phi + I_5 \cos \phi + \frac{D'}{D} I_6 = 0, \label{ceq1-iv} \\& &  I_1 I_8 - I_2 I_9 = 0. \label{ceq2-iv}
\end{eqnarray}
Here, the constants of motion representing conservation of energy, angular momentum and $z$ component of the momentum are $p_t = I_1, p_{\phi} = I_6$ and $p_z = I_2$, respectively. Thus, we can write the Eq. (\ref{r2-iv}) as
\begin{equation}\label{r22-iv}
\dot{r}^2 = p_t^2 - V(r),
\end{equation}
where the effective potential is defined by
\begin{equation}\label{Veff-iv}
V(r) = \frac{p_{\phi}^2}{D^2} + p_z^2 + \epsilon.
\end{equation}
Differentiating $\dot{r}$ given in the Eq. (\ref{r-iv}) with respect to proper time $\tau$ and using (\ref{phi-iv}) and (\ref{ceq1-iv}), we obtain
\begin{equation}\label{r222-iv}
\ddot{r} = p_{\phi}^2 \frac{D'}{D^3},
\end{equation}
which is again a Lienard type differential equation. By taking $\dot{D} = D' \dot{r}$, the  integration of (\ref{r222-iv}) with respect to $\tau$ is  reduced to the same form with (\ref{r2-iv}). Now, we introduce a new variable $\sigma = m^2 D^2 = \sinh^2 (m r)$ for $m^2 > 0$. Thus the Eq. (\ref{r22-iv}) in terms of the new variable yields
\begin{equation}\label{deq1-iv}
\dot{\sigma}^2 = 4 m^2 p_t^2 \left[ - \eta \sigma^2 - (\eta + m^2 \gamma^2) \sigma - m^2 \gamma^2 \right],
\end{equation}
where $\eta = (p_z^2 + \epsilon)/p_t^2 -1$ and $\gamma = p_{\phi} / p_t$. The general solution of this equation is
\begin{equation}\label{soln-deq1-iv}
\sigma (\tau) = \frac{1}{2 \eta} \left[ - ( \eta + m^2 \gamma^2) + (\eta - m^2 \gamma^2) \sin \left( 2 m p_t \sqrt{\eta} (\tau -\tau_0) \right) \right].
\end{equation}
Considering this new variable in the Eq. (\ref{phi-iv}) for $\phi (\tau)$ we get
\begin{equation}\label{phi2-iv}
\dot{\phi} = - \frac{m^2 p_{\phi}}{\sigma},
\end{equation}
which has the general solution as
\begin{equation}\label{soln-phi2-iv}
\phi (\tau) = - \arctan \left[  \frac{\eta - m^2 \gamma^2 -(\eta + m^2 \gamma^2) \tan \left( m p_t \sqrt{\eta} (\tau -\tau_0) \right)}{2 m \gamma \sqrt{\eta}} \right] + \phi_0.
\end{equation}
When the new variable $\sigma$ has the form $\sigma = \mu^2 D^2 = \sin^2 (\mu r)$ for $\mu^2 = -m^2 >0$, the Eq. (\ref{r22-iv}) give
\begin{equation}\label{deq2-iv}
\dot{\sigma}^2 = 4 \mu^2 p_t^2 \left[ \eta \sigma^2 - (\eta - \mu^2 \gamma^2) \sigma - \mu^2 \gamma^2 \right],
\end{equation}
and the general solution of this equation is
\begin{equation}\label{soln-deq2-iv}
\sigma(\tau) = \frac{1}{8 \eta^{3/2}} \left[ 4 \sqrt{\eta} (\eta - \mu^2 \gamma^2) + (\mu^2 \gamma^2 - \eta) e^{2 \mu p_t \sqrt{\eta} (\tau- \tau_0)}+ 4 \eta e^{- 2 \mu p_t \sqrt{\eta} (\tau- \tau_0)} \right].
\end{equation}
Then the Eq. (\ref{phi-iv}) for $\phi (\tau)$ has the general solution
\begin{equation}\label{soln-phi22-iv}
\phi (\tau) = \frac{ - 2 \mu \gamma  \sqrt{\eta}}{\sqrt{(\mu^2 \gamma^2 - \eta) (1+ \mu^2 \gamma^2 -\eta)}} \arctan \left[ \frac{1}{2} \sqrt{\frac{ (\mu^2 \gamma^2 -\eta) }{\eta (1+ \mu^2 \gamma^2 - \eta)}} \left( e^{2 \mu p_t \sqrt{\eta} (\tau- \tau_0)} - 2 \sqrt{\eta} \right) \right] + \phi_0,
\end{equation}
where $\mu^2 \gamma^2 >0$ and $ \eta (1 + \mu^2 \gamma^2 - \eta) >0$.

\section{Conclusions}
\label{conc}

In this paper, we have obtained the NGSs of geodesic Lagrangian $L$ for  G\"{o}del-type spacetimes for classes I, II, III and IV for which we have found {\it 7 NGS} generators. Thus, the G\"{o}del-type spacetimes corresponding to those classes admit the algebra $\mathcal{N}_7 \supset \mathcal{G}_5$. In special class I (where $m^2 = 4 w^2$) and class IV,  we have found {\it 9 NGS} generators. The NGS algebra admitted by the special class I is $\mathcal{N}_9 \supset \mathcal{G}_7$. The G\"{o}del-type spacetime in class IV admits the algebra $\mathcal{N}_9 \supset \mathcal{G}_6$.

We obtained the first integrals admitted by geodesic Lagrangians for the G\"{o}del-type spacetimes of each class I, II, III and IV, that are due to the existence of NGS vector fields including the KVs. Using the obtained first integrals in all classes of G\"{o}del-type spacetimes, we have derived the analytical solutions of geodesic equations which represents the usefulness of the NGSs. As stated in the literature \cite{calvao2,grave} that the radial equation of the geodesic motion is depend on the radial coordinate and its derivative, for example the Eq. (\ref{r2}) for class I. But, in each classes of G\"{o}del-type spacetimes  using the NGSs found in this study, we derive a  radial equation which depends only on angular coordinate $\phi$ (the Eq. (\ref{r1}) for class I; the Eq. (\ref{r1-cII}) for class II, and the Eq. (\ref{r-iv}) for class IV), and also a constraint equation (the Eq. (\ref{ceq-cI}) for class I; the Eq. (\ref{ceq-cII}) for class II, the Eq. (\ref{ceq1-iv}) for class IV). This is a new and unknown property of geodesic motions for G\"{o}del-type spacetimes.  The behaviour of the geodesics in G\"{o}del's universe has been extensively examined by several authors \cite{calvao2,grave,novello}, therefore, this behaviour is not a subject of this study.

\section*{Acknowledgements}

This work was supported by The Scientific Research Projects Coordination Unit of Akdeniz University (BAP). Project Number: 2013.01.115.003. The author would like to thank The Abdus Salam International Centre for Theoretical Physics (ICTP) for the financial grant of the successful conference "Symmetries, Differential Equations and Applications (SDEA-II)", National University of Sciences and Technology, School of Natural Sciences, Islamabad, Pakistan held in January 27-30, 2014.


\begin{thebibliography}{99}

\bibitem{godel} K. G\"odel, \emph{An example of a new type of cosmological solution of Einstein's field equations of gravitation}, \emph{Rev. Mod. Phys.}{\bf 21}(1949) 447.

\bibitem{kramer} H. Stephani, D. Kramer, M. A. H. MacCallum, C. Hoenselaers and E. Herlt,
{\it Exact Solutions of Einstein Field Equations}, Cambridge
University Press, (2003).

\bibitem{rayc} A. K. Raychaudhuri and S. N. Thakurta, \emph{Homogeneous space-times of G\"{o}del type}, \emph{Phys. Rev. D}{\bf 22} (1980) 802.

\bibitem{rebo1} M. J. Rebou\c{c}as and J. Tiomno, \emph{Homogeneity of Riemannian space-times of G\"{o}del type}, \emph{Phys. Rev. D}{\bf 28} (1983) 1251.

\bibitem{calvao1} M. O. Calv\~{a}o, M. J. Rebou\c{c}as, A.F.F Teixeira and W.M. Silva, \emph{Notes on a class of homogeneous space-times}, \emph{J. Math. Phys.} {\bf 29} (1988) 683.

\bibitem{teix} A. F. F. Teixeira, M. J. Rebou\c{c}as and J. E. \AA man, \emph{Isometries of homogeneous G\"{o}del type spacetimes}, \emph{Phys. Rev. D}{\bf 32} (1985) 3309.

\bibitem{rebo2} M. J. Rebou\c{c}as and J. Tiomno, \emph{A class of inhomogeneous G\"{o}del-type models},  \emph{Nuova Cimento B}{\bf 90} (1985) 204.

\bibitem{rebo3} M. J. Rebou\c{c}as and J. E. \AA man, \emph{Computer-aided study of a class of Riemannian space-times}, \emph{J. Math. Phys.} {\bf 28} (1987) 888.

\bibitem{adler} R. Adler, M. Bazin and M. Schiffer, \emph{Introduction to General Relativity}, McGraw-Hill, New York, 2nd Ed., (1975).

\bibitem{misner} C. W. Misner, K. S. Thorne and J. Wheeler, \emph{Gravitation}, W. H. Freeman, San Francisco, Ch.25, (1973).

\bibitem{ugur} U. Camci, \emph{Dirac analysis and integrability of geodesic equations for cylindrically symmetric spacetimes}, \emph{Int. J. Mod. Phys.}{\bf 12} (2003) 1431.


\bibitem{kundt} W. Kundt, \emph{Tr\"{a}gheitsbahnen in einem von G\"{o}del angegebenen Modell}, \emph{Z. Phys.}{\bf 145} (1956) 661.

\bibitem{chandra} S. Chandrasekhar and J.P. Wright, \emph{The geodesics in G\"{o}del's universe}, \emph{Proc.Natl.Acad.Sci.}{\bf 47} (1961) 341.

\bibitem{novello} M. Novello, I. D. Soares and J. Tiomno, \emph{Geodesic motion and confinement in G\"{o}del universe}, \emph{Phys. Rev. D}{\bf 27} (1983) 779.

\bibitem{rebo4} M. J. Rebou\c{c}as and A. F. F. Teixeira, \emph{Features of a relativistic space-time with seven isometries}, \emph{Phys. Rev. D}{\bf 34} (1986) 2985.

\bibitem{paiva} F. M. Paiva, M. J. Rebou\c{c}as and A.F.F Teixeira, \emph{Time-travel in the homogeneous Som-Raychaudhuri universe}, \emph{Phys.Lett.A}{\bf 126} (1987) 168.

\bibitem{som} M. M. Som and A. K. Raychaudhuri, \emph{Cylindrically Symmetric Charged Dust Distributions in Rigid Rotation in General Relativity}, \emph{Proc. Roy. Soc. London }{\bf A 304} (1968) 81.

\bibitem{calvao2} M. O. Calv\~{a}o, I.D. Soares and J. Tiomno, \emph{Geodesics in G\"{o}del-type space-times}, \emph{Gen. Rel. Grav.}{\bf 22} (1990) 683.

\bibitem{grave} F. Grave, M. Bauser, T. M\"{u}ller, G. Wunner and W.P. Schleich, \emph{The G\"{o}del universe:Exact geometrical optics and analytical investigation on motion}, \emph{Phys. Rev. D}{\bf 80} (2009) 103002.

\bibitem{kajari} E. Kjari, R. Walseri W. P. Schleich and A. Delgado, "Sagnac effect of G\"{o}del's universe", \emph{Gen. Rel. Grav.}{\bf 36} (2004) 2289.

\bibitem{daut} G. Dautcourt, \emph{The lightcone of G\"{o}del-like spacetimes}, \emph{Class. Quant. Grav.}{\bf 27} (2010) 225024.


\bibitem{katzin} G. H. Katzin, J. Levine and W. R. Davis, \emph{Curvature collineations: A fundamental symmetry property of space-times of general relativity defined by the vanishing Lie derivative of the Riemann curvature tensor}, \emph{J. Math. Phys.} {\bf 10} (1969) 617.

\bibitem{hall-costa} G. S. Hall and J. da Costa, \emph{Affine collineations in space-times}, \emph{J. Math. Phys.}\, {\bf 29} (1988) 2645.

\bibitem{melfo} A. Melfo, L. A. Nunez, U. Percoco and V. M. Villalba, \emph{Collineations of G\"{o}del-type space-times}, \emph{J. Math. Phys.}\,{\bf 33} (1992) 2258.

\bibitem{tsamparlis} M. Tsamparlis, D. Nikolopoulos and P. S. Apostolopoulos, \emph{Computation of the conformal algebra of 1+3 decomposable spacetimes}, \emph{Class. Quant. Grav.}\, {\bf 15} (1998) 2909.

\bibitem{cmc-sharif} U. Camci and M. Sharif, \emph{matter collineations of spacetime homogeneous G\"{o}del-type metrics}, \emph{Class. Quant. Grav.} {\bf 20} (2003) 2169.

\bibitem{capo93} S. Capozziello and R. de Ritis, \emph{Relation between the potential and nonminimal coupling in inflationary cosmology}, \emph{Phys. Lett. A} {\bf 177} (1993) 1.

\bibitem{capo2008} S. Capozziello and A. de Felice, \emph{f(R) cosmology from Noether's symmetry}, \emph{JCAP}{\bf 08} (2008) 016.

\bibitem{capo2009} S. Capozziello, E. Piedipalumbo, C. Rubano and P. Scudellaro, \emph{Noether symmetry approach in phantom quintessence}, \emph{Phys. Rev. D} {\bf 80} (2009) 104030.

\bibitem{capo94} S. Capozziello and R. de Ritis, \emph{Spherically symmetric solutions in f(R) gravity via the Noether symmetry approach}, \emph{Class. Quant. Grav.}{\bf 24} (1994) 2153.


\bibitem{sanyal01} A. K. Sanyal and B. Modak, \emph{Is Noether symmetric approach consistent with dynamical equation in nonminimal scalar tensor theories?}, {\it Class. Quant. Grav.}{\bf 18} (2001) 3767.

\bibitem{sanyal02} A. K. Sanyal, \emph{Noether and some other dynamical symmetries in Kantowski-Sachs model}, \emph{Phys. Lett. B}{\bf 524} (2002) 177.

\bibitem{camci} U. Camci and Y. Kucukakca, \emph{Noether symmetries of Bianchi I, Bianchi III and Kantowski-Sachs spacetimes in scalar-coupled gravity theories}, \emph{Phys. Rev. D}{\bf 76} (2007) 084023.

\bibitem{camci3} Y. Kucukakca, U. Camci and I. Semiz, \emph{LRS Bianchi type-I universes exhibiting Noether symmetry in the scalar-tensor Brans-Dicke theory}, \emph{Gen. Rel. Grav.}\,{\bf 44} (2012) 1893.

\bibitem{sharif2013} M. Sharif and S. Waheed, \emph{Noether symmetries of some homogeneous universe models in curvature corrected scalar-tensor gravity}, \emph{JCAP}{\bf 02} (2013) 043.

\bibitem{feroze1} T. Feroze, F.M. Mahomed and A. Qadir, \emph{The connection between isometries and symmetries of geodesic equations of underlying spaces}, \emph{Nonlinear Dynam.}{\bf 45} (2006) 65.

\bibitem{feroze2} T. Feroze, \emph{New conserved quantities for the spaces of different curvatures}, \emph{Modern Phys.Lett.}{\bf A25} (2010) 1107.

\bibitem{feroze3} T. Feroze and I. Hussain, \emph{Noether symmetries and conserved quantities for spaces with a section of zero curvature}, \emph{J. Geom. Phys.} {\bf 61} (2011) 658.

\bibitem{tsamparlis1} M. Tsamparlis and A. Paliathanasis, \emph{Lie and Noether symmetries of geodesic equations and collineations}, \emph{Gen.Rel.Grav.}{\bf 42} (2010) 2957.

\bibitem{tsamparlis2} M. Tsamparlis and A. Paliathanasis, \emph{The geometric nature of Lie and Noether symmetries}, \emph{Gen.Rel.Grav.}{\bf 43} (2011) 1861.

\bibitem{feroze4} F. Ali and T. Feroze, \emph{Classification of plane symmetric static space-times according to their Noether symmetries}, \emph{Int. J. Theor. Phys.}{\bf 52} (2013) 3329.

\bibitem{camci2} Y. Kucukakca and U. Camci, \emph{Noether gauge symmetry for f(R) gravity in Palatini formalism}, \emph{Astrophys. Space Sci.}{\bf 338} (2011) 211.

\bibitem{jamil} M. Jamil, F. M. Mahomed and D. Momeni, \emph{Noether symmetry approach in f(R) tachyon model}, \emph{Phys. Lett. B}{\bf 702} (2011) 315.

\bibitem{ibrar} I. Hussain, M. Jamil and F. M. Mahomed, \emph{Noether gauge symmetry approach in f(R) gravity}, \emph{Astrophys. Space Sci.}{\bf 337} (2012) 339.

\bibitem{yusuf} Y. Kucukakca, \emph{Scalar tensor teleparallel dark gravity via Noether Symmetry}, \emph{Eur. Phys. J. C} {\bf 73} (2013) 2327.

\bibitem{capo2012} S. Cpozziello, M. De Laurentis and S. D. Odintsov, \emph{Hamiltonian dynamics and Noether Symmetries in Extended gravity Cosmology}, \emph{Eur. Phys. J. C} {\bf 72} (2012) 2068.

\bibitem{basilakos} S. Basilakos, M. Tsamparlis and A. Paliathanasis, \emph{Using the Noether symmetry approaqch to probe the nature of dark enegy}, \emph{Phys. Rev. D}{\bf 83} (2011) 103512.

\bibitem{noether} E. Noether, \emph{Invariante Variationsprobleme}, \emph{G\"ottingen Math. Phys. Kl.} {\bf 2} (1918) 235; English translation by M.A. Tavel,\emph{Invariant Variation Problems}, \emph{Transport Theory and Statistical Physics}\,{\bf 1(3)} (1971) 186.

\bibitem{polyanin} A. D. Polyanin and V. F. Zaitsev, \emph{Handbook of Exact Solutions for Ordinary Differential Equations}, 2nd Ed., Chapman and Hall/CRC Press, (2002).


\end{thebibliography}
\end{document}